\documentclass[12pt]{article}
\usepackage{latexsym}
\usepackage{epsfig}

\newcommand{\beq}{\begin{eqnarray}}
\newcommand{\eeq}{\end{eqnarray}}
\newcommand{\be}{\begin{eqnarray*}}
\newcommand{\ee}{\end{eqnarray*}}
\newcommand{\bk}{{\bf k}}
\newcommand{\bp}{{\bf p}}

\newcommand{\ve}{\varepsilon}

\newcommand{\om}{{\omega}}

\begin{document}

\centerline{\Large\bf{On the Casimir effect in a continuous medium}}

\bigskip
\centerline{Finn Ravndal}
\bigskip
\centerline{\it Department of Physics, University of Oslo, Blindern, N-0316 Oslo, Norway.}

\begin{abstract}

\small{It is pointed out that the Casimir energy in a medium can be obtained most directly from the zero-point energy of the electromagnetic field because of its reduced propagation velocity. This brings to the fore again the old problem related to how the principle of relativity is combined with the Maxwell field equations in a continuous medium.}

\end{abstract}

In a recent paper Brevik and Milton\cite{BM} calculates the Casimir force between two parallel, metallic plates separated by a dielectric medium with index of refraction $n$. It is derived from the Minkowski energy-momentum tensor\cite{Minkowski}  where the field correlators are obtained from the fluctuation-dissipation theorem combined with more standard Green's functions methods. After a rather lengthy calculation and neglecting the effects of non-linear dispersion, they obtain simply the ordinary vacuum result reduced by the factor $n$.

Such a simple result asks for a more direct derivation.  In fact, that is possible by the alternative and perhaps more common method of deriving it from the energy of the electromagnetic zero-point fluctuations between the plates.  Since an electromagnetic wave in the medium propagates with the velocity $c =1/n$ when the velocity in vacuum is set equal to $c_0=1$, a photon with wave vector $\bk$ will then have the energy  $\ve_\bk = \hbar\om_\bk$  where the frequency $\om_\bk = ck = |\bk|/n$. The zero-point energy is again given by the standard expression $\sum_\bk \hbar\om_\bk = (\hbar/n)\sum_\bk|\bk|$. Except for the factor $1/n$, this  gives just the standard Casimir energy between the plates after regularization. We thus have reproduced their result without any calculations.

This more direct derivation should not come as a surprise since the physics on which it is based, is consistent with the Minkowski theory in the rest frame of the medium where also Brevik and Milton perform their calculation. But it has been known for exactly one hundred years that this theory has the basic problem that the resulting energy-momentum tensor is not symmetric as it should be\cite{eff}.  It has recently been stressed that this problem is related to the basic property that the theory was made to  be valid in any inertial frame moving with respect to the rest frame, i.e. invariant under vacuum Lorentz transformations\cite{FR}.  To the author of this alternative theory, it is not obvious that this mathematical requirement is necessary from a physical point of view.

As a direct consequence of this requirement, the mass-squared $\ve^2 - p^2$ of a photon with momentum $\bp = \hbar\bk$, is negative. It is therefore some kind of tachyon. As stated by Brevik and Milton\cite{BM} in the beginning of Chapter 4, this theory now makes it possible to explain the Cerenkov radiation from an electron moving through the medium by going to the rest frame of the incoming electron. Here it can decay into a new electron together with a photon with negative energy moving in the opposite direction of the final electron. Needless to say, such photons are very different from what we usually mean with that name.

In the alternative theory\cite{FR} one abandons the requirement of vacuum Lorentz invariance and restricts its use to the rest frame of the medium. The electromagnetic Lagrangian is then invariant under Lorentz transformations based on the physical light velocity $1/n$ in the medium. For free fields it is the same as used by Glauber and Lewenstein in their investigation of electromagnetic field fluctuations in media\cite{GL}. One can now longer describe a phenomenon like the Cerenkov effect in the rest frame of the particle moving through the medium. The invariant squared photon mass is then $n^2\ve^2 - p^2 = 0$ and the photon can therefore be said to be massless. This is exactly the same as for other, similar theories describing excitations with linear dispersion relations in condensed matter physics.

We want to thank Yuri Galperin and Bo-Sture Skagerstam for useful discussions.

\end{document}